# Post-Quantum Cryptography: A Zero-Knowledge Authentication Protocol

J.P. Hecht[1]

[1]Universidad de Buenos Aires, Facultades de Ciencias Económicas, Ciencias Exactas y Naturales e Ingeniería, Maestría en Seguridad Informática, Buenos Aires, Argentina
[1] qubit101@gmail.com

*Abstract-* **In this paper, we present a simple bare-bones solution of a Zero-Knowledge authentication protocol which uses non-commutative algebra and a variation of the generalized symmetric decomposition problem (GSDP) as a one-way function. The cryptographic security is assured as long the GSDP problem is computationally hard to solve in non-commutative algebraic structures and belongs currently to the PQC category as no quantum computer attack is likely to exists.**

## I. INTRODUCTION

Post-quantum cryptography (PQC) has achieved an official NIST (USA) status [1][2] and its principal purpose is to find cryptographic protocols that resist quantum attacks like Shor's algorithm [3], which theoretically solves some one-way trap functions in polynomial time like the integer factorization problem (IFP) and the discrete logarithm problem (DLP) in numerical fields. But this is not the only reason behind the development of this new kind of solutions, they pretend to defend against recent developments of quasi-polynomic algorithms for solving the low characteristic discrete logarithm problem [4] and to prevent attacks against pseudorandom bit generators [5], vastly used by current numeric field based algorithms of asymmetric cryptography.

Since the beginning of the past decade, a great number of post-quantum proposals were formulated [6]. Among them, many non-commutative and non-associative algebraic solutions stand out [7][8][9][10][11][12][13][14][15] [16] [17] [18][19][20][21][22][23][24].

Zero-Knowledge authentication is fully described in references like [25][26][27][28][29][30], and no attempt is made here to explain such details. Our purpose is simply to present a plain sketch of a potential PQC solution.

## II. GENERAL DESCRIPTION

The protocol works with the general linear group [31] $GL(d, F_p)$, were $d \in \mathbb{Z}$ and $p$ is a prime. Being a non-commutative group, it is mandatory to find commutative substructures inside to develop asymmetric protocols. There are, beside others, three simple ways to achieve it [32].

We use here the second way, that is the fact that two matrices commute *iff* they share the same orthogonal basis [33].

Suppose we start with two full ranked diagonal matrices and conjugate both with any non-singular same order square matrix; then those similar matrices commute.

As an example, we choose $d=8$ and $F_{251}$ operations. An extensive description of $GL(8, F_{251})$ is given in [32].

## III. GSDP PROBLEM

The Generalized Symmetric Decomposition Problem (GSDP) could be stated [13] as

$$Given\ G\ a\ non-commutative\ group\ and\ S\ a\ (hidden)\ commutative\ subgroup, knowing\ (x,y) \in G^2\ and\ (m,n) \in \mathbb{Z}^2,\ find\ z \in S\ where\ y = z^m\ x\ z^n \quad (1)$$

Neither a polynomic-time solution for (1) is currently known nor any quantum computing based attack seems feasible is likely to exist. Of course, this statement is only a conjecture and is far from being proved. On the other side, no currently proposed and accepted PQC solution [1], has attained today a mathematical proved status.

## IV. ZKP PROTOCOL

*STEP 0 – AGREEMENT*
A community of entities (individually called *provers* and *verifiers*, where roles could be changed at will) agrees over the use the general linear group $GL(8, F_{251})$. Definitions of $M_8$, $P_8$ and other symbols are stated in [32]. Besides they agree about the following public parameters:

$$\begin{aligned}(P,G) &\in_R M_8^2 \\ (m,n) &\in_R \mathbb{Z}^{+2}\end{aligned} \quad (2)$$

A reasonable upper limit for random integers in that space ($d=8$, $p=251$) could be $65536$. It is of interest that each selected random matrix should have a high multiplicative order, a feature associated with the irreducibility (or better primitivity) of the characteristic polynomial in the simple algebraic extension $F_{251}[x]$ [31].

*STEP 1 – PRIVATE AND PUBLIC KEYS*
Each entity *(Alice, Bob , …)* should define a private key generating a random diagonal matrix with non-repeating values chosen in $\mathbb{Z}_{251}^*$.

$$Select\ unique\ valued\ set\ (\lambda_1 \ldots \lambda_8) \in_R \mathbb{Z}_{251}^* \quad (3)$$

And with each diagonal matrix obtained $(D_A, D_B, \ldots)$ generate the private key:

$$\begin{aligned}(A, B, \ldots)\ private\ keys \in P_8 \\ A = P D_A P^{-1}, B = P D_B P^{-1}, \ldots \end{aligned} \quad (4)$$

Clearly private keys and their powers commute. Now, using one-way GSDP the corresponding public keys are derived.

$$\begin{aligned}(G_A, G_B, \ldots)\ public\ keys \in M_8 \\ G_A = A^m G\ A^n, G_B = B^m G\ B^n, \ldots \end{aligned} \quad (5)$$

*STEP 2 – WITNESS (Alice as prover, Bob as verifier)*
Alice generates the witness $S$ and sends it to Bob.

$$\begin{aligned} k \in_R \mathbb{Z}^+ \\ S = A^k G_B\ A^{-m} \implies \end{aligned} \quad (6)$$

*STEP 3 – CHALLENGE*
Bob generates challenge bit $b$ and question Q, and send both to Alice.
$$\begin{aligned} b \in_R \{0,1\} \implies \\ if\ b = 0\ then\ H \in_R M_8\ and\ Q = B^m H\ B^n \implies \\ if\ b = 1\ then\ Q = B^m S\ G_A\ B^n \implies \end{aligned} \quad (7)$$

*STEP 4 – RESPONSE*
Alice generates response $R$ and sends it to Bob.

$$\begin{aligned} if\ b = 0\ then\ R = S^{-m} Q\ S^{-n} \implies \\ if\ b = 1\ then\ R = A^{-k} Q\ A^{-n} \implies \end{aligned} \quad (8)$$

*STEP 5 – IDENTITY VALIDATION*
Bob verifies response $R$ and accepts or rejects Alice identity.
$$\begin{aligned} if\ b = 0\ accept\ if\ Q = S^m R\ S^n \\ if\ b = 1\ accept\ if\ G_B G = B^{-m} R\ B^{-n} \end{aligned} \quad (9)$$

If rejection occurs in the last step, Bob forces the repetition of steps 2 to 5 until he is fully satisfied with Alice's identity. Else he rejects the prover's identity. Observe that no secret keys are revealed unless GSDP is solved. Validation is justified according to

$$if\ b = 0\ then\ S^m R\ S^n = S^m S^{-m} Q\ S^{-n} S^n = Q$$
$$if\ b = 1\ then$$
$$B^{-m} R\ B^{-n} = B^{-m} A^{-k} Q\ A^{-n}\ B^{-n} =$$
$$= B^{-m} A^{-k} B^m S\ G_A\ B^n\ A^{-n}\ B^{-n} =$$
$$= B^{-m} A^{-k} B^m A^k\ G_B\ A^{-m}\ A^m G\ A^n\ B^n\ A^{-n}\ B^{-n} =$$
$$= G_B G \quad (10)$$

If $b=0$, to impersonate Alice no private key is needed, as the fake Alice forges any $S^*$ that Bob validates. But if $b=1$, it is mandatory to use the private key of Alice to be validated. It would be a bad strategy for Bob to replace a random bit with $b=1$ in any round, as any entity could consistently impersonate Alice using a slight variation of the simulator algorithm [25][28], which is explained in the next section.

V. ZKP CONDITION

An interactive authentication protocol fulfills ZKP condition *iff* it complies with three properties: *Completeness*, *Soundness* and *Zero-Knowledge* [25][28].

The first condition could be proven considering that any entity possessing Alice's private key could be verified by Bob, whatever challenges he receives. So, any honest verifier will accept Alice's identity. Of course, only the true Alice would be in condition of always providing right answers.

The second condition implies that a dishonest prover and an honest verifier will fail in half of the rounds, specifically when he receives a $b=1$ challenge. Suppose Mallory is such an entity, he invents a random private key $A^*$ belonging to $P_8$. If he receives $b=0$ as a challenge, it suffices to generate a random $S^*$ witness, as said before. But if $b=1$ is the challenge, he works out a fake witness and responses $S^*=A^{*k} G_B A^{*-m}$ and $R^*= A^{*-k} Q A^{*-n}$. But this will not fulfill the validation as

$$B^{-m} R^* B^{-n} = B^{-m} A^{*-k} Q\ A^{*-n}\ B^{-n} =$$
$$= B^{-m} A^{*-k} B^m S\ G_A\ B^n\ A^{*-n}\ B^{-n} =$$
$$= B^{-m} A^{*-k} B^m A^{*k}\ G_B\ A^{*-m}\ A^m G\ A^n\ B^n\ A^{*-n}\ B^{-n} =$$
$$= G_B (A^{*-m}\ A^m) G\ (A^n A^{*-n}) \neq G_B G \quad (11)$$

Clearly the public key of Alice does not match or simplify the faked private key of Mallory.

The third condition states that the protocol has the ZKP property *iff* there exists a simulator algorithm [25][28] that mimics valid session reports with faked values, fully undistinguishable from the true ones. A simulator algorithm is presented here.

1. **SELECT CHALLENGE BIT AND GENERATE A WITNESS**
   *Make random bit b $\varepsilon_R$ [0, 1]*
   *If b=0 → Generate random witness S=*S $\varepsilon_R$ $M_8$*
   *If b=1 → Generate witness S=$G_B$ G $G_A^{-1}$*
2. **RECEIVE CHALLENGE**
   *If b=0 → Generate random *Q $\varepsilon_R$ $M_8$*
   *If b=1 → Q=$B^m S G_A B^n$=$B^m G_B G(G_A^{-1} G_A)B^n$ = $B^m G_B G B^n$*
3. **GIVE RESPONSE**
   *If b=0 → *R=*$S^{-m}$*Q *$S^{-n}$*
   *If b=1 → R=Q*
4. **VERIFY AND REGISTER SESSION**
   *If b=0 → *$S^m$ *R* $S^n$=*Q*
   *If b=1 → $B^{-m}RB^{-n}$=($B^{-m}B^m$) $G_B G(B^n B^{-n})=G_B G$*
   *Append in order the 4-tuple (S, Q, b, R) to the sesión log..*
5. **ITERATE 1. TO 4. MANY TIMES**

Clearly no verifier would accept the identity extracted from the simulation protocol as he does not generate the random challenge and otherwise it would verify only half of the round sessions.

## VI. Security

Because public keys, witness, challenge question and response are GSDP strong protected, a natural way to defeat this protocol would be a brute-force attack over the private keys space. The cardinal involved is:

$$|P_8| = 249.248.247.246.245.244.243.242 = \\ = 13190481178699144320 \approx 10^{19} \approx 2^{64} \quad (12)$$

This implies for ($d=8$, $p=251$) a security of 64 bits, and this level could be easily increased using higher dimensions. For example, ($d=16$, $p=251$) provides a security of 127 bits. Other details could be found at [29][30][32].

## VII. Conclusion

It is presented here a sketch of a ZKP protocol using a non-commutative algebraic structure and $Z_{251}$ arithmetic. In the presented protocol, the security and the hiding of the private elements relies on the GSDP one way function, which belongs, as conjectured, to the PQC family.

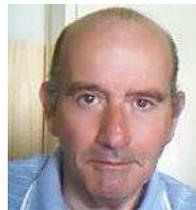

**Pedro Hecht** received an MSci in Information Technology at Escuela Superior de Investigación Operativa and an PhD degree from Universidad de Buenos Aires (UBA). Currently, he is full professor of cryptography at Information Security Graduate School at UBA, EST (Army Engineering School) and IUPFA (Federal Police University), he is research fellow UBACyT and Director of EUDEBA editorial board of UBA.